\PassOptionsToPackage{unicode}{hyperref}
\PassOptionsToPackage{hyphens}{url}
\documentclass[
  11pt]{article}
\usepackage{xcolor}
\usepackage[margin=1in]{geometry}
\usepackage{amsmath,amssymb}
\usepackage{iftex}
\ifPDFTeX
  \usepackage[T1]{fontenc}
  \usepackage[utf8]{inputenc}
  \usepackage{textcomp} 
\else 
  \usepackage{unicode-math} 
  \defaultfontfeatures{Scale=MatchLowercase}
  \defaultfontfeatures[\rmfamily]{Ligatures=TeX,Scale=1}
\fi
\usepackage{lmodern}
\ifPDFTeX\else
\fi
\IfFileExists{upquote.sty}{\usepackage{upquote}}{}
\IfFileExists{microtype.sty}{
  \usepackage[]{microtype}
  \UseMicrotypeSet[protrusion]{basicmath} 
}{}
\makeatletter
\@ifundefined{KOMAClassName}{
  \IfFileExists{parskip.sty}{%
    \usepackage{parskip}
  }{
    \setlength{\parindent}{0pt}
    \setlength{\parskip}{6pt plus 2pt minus 1pt}}
}{
  \KOMAoptions{parskip=half}}
\makeatother
\usepackage{longtable,booktabs,array}
\usepackage{calc} 
\usepackage{etoolbox}
\makeatletter
\patchcmd\longtable{\par}{\if@noskipsec\mbox{}\fi\par}{}{}
\makeatother
\IfFileExists{footnotehyper.sty}{\usepackage{footnotehyper}}{\usepackage{footnote}}
\makesavenoteenv{longtable}
\usepackage{bookmark}
\IfFileExists{xurl.sty}{\usepackage{xurl}}{} 
\hypersetup{
  hidelinks,
  pdfcreator={LaTeX via pandoc}}

\author{Jayson Steffens\\\texttt{github.com/stffns/vstash}}
\date{}

\begin{document}

\section{vstash: Local-First Hybrid Retrieval with Adaptive Fusion for LLM Agents}\label{vstash-local-first-hybrid-retrieval-with-adaptive-fusion-for-llm-agents}

\textbf{Jayson Steffens} \href{https://github.com/stffns/vstash}{github.com/stffns/vstash}

\begin{center}\rule{0.5\linewidth}{0.5pt}\end{center}

\subsection{Abstract}\label{abstract}

We present \textbf{vstash}, a local-first document memory system that combines vector similarity search with full-text keyword matching via Reciprocal Rank Fusion (RRF) and adaptive per-query IDF weighting. All data resides in a single SQLite file using sqlite-vec for approximate nearest neighbor search and FTS5 for keyword matching.

We make four primary contributions. \textbf{(1)} Self-supervised embedding refinement via hybrid retrieval disagreement: across 753 BEIR queries on SciFact, NFCorpus, and FiQA, 74.5\% produce top-10 disagreement between vector-heavy (vec=0.95, fts=0.05) and FTS-heavy (vec=0.05, fts=0.95) search (per-dataset rates 63.4\% / 73.4\% / 86.7\%, Section 5.2), providing a free training signal without human labels. Fine-tuning BGE-small (33M params) with MultipleNegativesRankingLoss on 76K disagreement triples improves NDCG@10 on all 5 BEIR datasets (up to +19.5\% on NFCorpus vs.~BGE-small base RRF, Table 6). On 3 of 5 datasets, under different preprocessing, the tuned 33M-parameter pipeline matches or exceeds published ColBERTv2 results (110M params) and an untrained BGE-base (110M); on FiQA and ArguAna it underperforms ColBERTv2 (Section 5.5). \textbf{(2)} Adaptive RRF with per-query IDF weighting improves NDCG@10 on all 5 BEIR datasets versus fixed weights (up to +21.4\% on ArguAna), achieving 0.7263 on SciFact with BGE-small. \textbf{(3)} A negative result on post-RRF scoring: frequency+decay, history-augmented recall, and cross-encoder reranking all failed to improve NDCG. \textbf{(4)} A production-grade substrate with integrity checking, schema versioning, ranking diagnostics, and a distance-based relevance signal validated on 50,425 relevance-judged queries across the 5 BEIR datasets.

Search latency remains 20.9 ms median at 50K chunks with stable NDCG. The fine-tuned model is published as \texttt{Stffens/bge-small-rrf-v2} on HuggingFace. All code, data, and experiments are open-source.

\begin{center}\rule{0.5\linewidth}{0.5pt}\end{center}

\subsection{1 Introduction}\label{introduction}

Large language model agents increasingly require persistent memory -- the ability to store, retrieve, and prioritize information across sessions. While cloud-hosted vector databases serve this need at scale, many use cases demand local-first operation: developer tooling, personal knowledge management, privacy-sensitive workflows, and offline agents.

Existing local solutions face three gaps:

\textbf{Retrieval quality.} Pure vector search misses exact keywords (error codes, proper names); pure keyword search misses semantic paraphrases. Hybrid fusion helps, but combining scores from incompatible distributions (cosine distance vs.~BM25 rank) is non-trivial.

\textbf{Temporal awareness.} Documents accessed yesterday should rank differently from documents untouched for months. Most RAG systems treat all chunks equally regardless of usage history.

\textbf{Confidence estimation.} When every query returns results with uniformly high scores, the system cannot distinguish ``I found something relevant'' from ``I returned the least irrelevant thing I have.''

We introduce \textbf{vstash}, a single-file system built on SQLite that addresses all three gaps. Our key insight is that adaptive RRF fusion -- adjusting vector and keyword weights per query using IDF analysis -- combined with MMR deduplication and distance-based relevance signaling, produces retrieval quality competitive with published baselines on standard benchmarks.

\subsubsection{Contributions}\label{contributions}

Our primary contributions are: (1) a self-supervised embedding refinement method exploiting vector/FTS disagreement, yielding a 33M-parameter pipeline that achieves NDCG comparable to or exceeding published ColBERTv2 results on 3 of 5 BEIR datasets with zero human labels, under different preprocessing conditions (Section 7.3); (2) adaptive RRF with IDF-based per-query weight adjustment, validated on 5 BEIR datasets; (3) a documented negative result showing that post-RRF scoring strategies (frequency+decay, history-augmented recall, cross-encoder reranking) fail to improve retrieval quality; and (4) a production-grade substrate with integrity checking, schema versioning, and ranking diagnostics.

Secondary contributions include intra-document MMR deduplication (+1.8\% NDCG@5), context expansion (2.64x richer LLM context at +0.12 ms), a distance-based relevance signal (F1 = 0.856-0.996 cross-domain on BEIR), hybrid code-aware chunking with a 3-tier splitting pipeline, and a \texttt{vstash\ retrain} CLI command that enables users to fine-tune embeddings on their own corpora using the same disagreement signal.

\begin{center}\rule{0.5\linewidth}{0.5pt}\end{center}

\subsection{2 Related Work}\label{related-work}

\textbf{Memory for LLM agents.} MemGPT (Packer et al., 2023) introduced virtual context management with explicit memory tiers. Mem0 (Chhikara et al., 2025) and Memoria (Sarin et al., 2025) provide production-ready memory layers with cloud backends. A-MEM (Xu et al., 2025) uses agentic self-organization. Unlike these systems, vstash operates entirely locally with zero cloud dependencies.

\textbf{Local-first agent memory.} Several concurrent projects share vstash's SQLite + hybrid search architecture: palinode (git-native markdown + sqlite-vec + RRF), cpersona (MCP server with 3-strategy RRF), agentmem (FTS5 + vector with adaptive ranking), and sqlite-memory (FTS5 + vector with offline sync). None publish formal NDCG evaluations on standard benchmarks. neo4j-labs/agent-memory takes an orthogonal approach with graph-native entity extraction and temporal knowledge graphs, targeting relational queries rather than dense retrieval.

\textbf{Hybrid retrieval.} Reciprocal Rank Fusion (Cormack et al., 2009) merges ranked lists without requiring comparable scores. Ma et al.~(2024) showed RRF outperforms learned re-rankers on out-of-domain data. Our contribution is adaptive per-query weight adjustment using IDF analysis, not explored in prior RRF work.

\textbf{Hard negative mining for dense retrieval.} The quality of dense retrieval models depends critically on the training signal. DPR (Karpukhin et al., 2020) established the use of BM25 negatives for training dense retrievers. ANCE (Xiong et al., 2020) introduced asynchronous hard negative mining from the model's own approximate nearest neighbors. STAR (Zhan et al., 2021) refined this with a two-stage approach using relevance labels as supervision. More recently, NV-Retriever (Lee et al., 2024) and BGE-M3 (Chen et al., 2024) demonstrated that carefully curated hard negatives at scale produce state-of-the-art results on MTEB and BEIR. Our approach (Section 5) differs from this lineage in two respects: the training signal comes from disagreement between two retrieval modalities (dense and sparse) within the same system rather than from external relevance labels or model-internal mining, and the triples are generated at zero cost as a byproduct of normal hybrid retrieval operation. The closest precedent is the use of BM25 negatives in DPR, but our signal is bidirectional -- capturing both dense blind spots (chunks found by FTS but missed by vector) and lexical blind spots (the reverse) -- whereas DPR mines only in one direction.

\textbf{Temporal decay in memory.} The Ebbinghaus forgetting curve (1885) inspires exponential decay models. Zep (2025) uses temporal knowledge graphs; MaRS (2025) models cognitive forgetting. We explored decay directly in the scoring formula but found it did not improve retrieval quality on benchmarks (Section 6).

\begin{center}\rule{0.5\linewidth}{0.5pt}\end{center}

\subsection{3 System Architecture}\label{system-architecture}

vstash stores all data in a single SQLite database using WAL mode for concurrent read safety. The database contains five core tables: \emph{documents} (metadata, hierarchical tags, source type), \emph{chunks} (text segments with sequence numbers and access counters), \emph{vec\_chunks} (sqlite-vec virtual table for ANN search, 384-dim float vectors), \emph{fts\_chunks} (FTS5 virtual table with Porter stemming), and \emph{journal\_entries} (append-only cross-session memory for agents).

\begin{verbatim}
                        INGESTION
  PDF/DOCX/URL/Code --> MarkItDown --> Chunking --> FastEmbed
                          parse      (3-tier code    (ONNX,
                                      or semantic)   384-dim)
                               |
                               v
                    SQLite (WAL mode)
  documents | chunks | vec_chunks (ANN) | fts_chunks (FTS5)
  journal_entries | profiles (multi-DB isolation)
                               |
                               v
                        RETRIEVAL
  Query --> Embed --+--> Vector ANN --+
                    |                 +--> RRF Fusion
                    +--> FTS5 BM25 ---+        |
                                    Recency Boost (opt-in) -->
                                    MMR Dedup -->
                                    Distance Signal -->
                                    Context Expansion (+/-1)
                               |
                               v
   CLI  |  Python SDK  |  MCP Server (16 tools)  |  Claude Hook
\end{verbatim}

\emph{Figure 1: vstash architecture.}

\subsubsection{3.1 Ingestion}\label{ingestion}

Documents are parsed via MarkItDown (PDF, DOCX, HTML, URLs) or read directly (code files). Text is split into chunks using semantic chunking or code-aware chunking (Appendix B) depending on source type. Chunks are embedded using FastEmbed (ONNX Runtime, BAAI/bge-small-en-v1.5) at approximately 700 chunks per second on CPU. Batch ingestion via \texttt{add\_documents\_batch()} amortizes transaction overhead for bulk loading.

\subsubsection{3.2 Hybrid Search with RRF}\label{hybrid-search-with-rrf}

Given a query \emph{q}, we retrieve candidates from both indexes and fuse via Reciprocal Rank Fusion:

\begin{verbatim}
RRF(c) = w_v / (k + r_v(c)) + w_f / (k + r_f(c))
\end{verbatim}

where \emph{r\_v(c)} and \emph{r\_f(c)} are the ranks of chunk \emph{c} in vector and FTS5 results respectively, \emph{k = 60}, and default weights \emph{w\_v = 0.6}, \emph{w\_f = 0.4}. These weights are overridden by adaptive IDF weighting (Section 4). Each query word is individually double-quoted and joined with OR, preventing FTS5 Boolean operator injection.

\subsubsection{3.3 Intra-Document MMR Deduplication}\label{intra-document-mmr-deduplication}

After RRF ranking, multiple chunks from the same document often cluster in the top-\emph{k}. We apply intra-document Maximal Marginal Relevance:

\begin{verbatim}
MMR(c) = L * norm_score(c) - (1 - L) * max_{s in S_d} cos_sim(emb(c), emb(s))
\end{verbatim}

where \emph{S\_d} is the set of already-selected chunks from the same document \emph{d}, and \emph{L = 0.5}. Chunks from different documents compete purely on score. When the best remaining candidate has negative MMR, selection stops. This improves diversity from approximately 3.2 to 5.0 unique documents per top-5 while improving NDCG@5 from 0.814 to 0.829 (+1.8\%).

\subsubsection{3.4 Context Expansion}\label{context-expansion}

A single chunk (approximately 250 tokens) provides insufficient context for LLM answer generation. We expand each search result by fetching adjacent chunks (+/-1 by sequence number within the same document). Default window \emph{w = 1} yields 2.64x more text per result at +0.12 ms overhead.

\subsubsection{3.5 Interfaces}\label{interfaces}

vstash integrates with LLM agents through three interfaces: a 16-tool MCP server, a Claude Code hook for automatic context injection on knowledge questions, and a Python SDK with context manager protocol. Details are provided in Appendix D.

\subsubsection{3.6 Relevance Signal via Vector Distance}\label{relevance-signal-via-vector-distance}

A retrieval system that always returns results -- even for off-topic queries -- must provide a confidence signal. We evaluated score spread (max - min of top-\emph{k} scores) and the cosine distance of the best vector match. Score spread requires scoring warm-up and produces complete class overlap (F1 = 0.667); vector distance works from the first search with zero class overlap (F1 = 0.952).

\textbf{Table 1: Relevance signal comparison -- pilot study (10 relevant + 10 irrelevant queries)}

{\def\LTcaptype{none} 
\begin{longtable}[]{@{}
  >{\raggedright\arraybackslash}p{(\linewidth - 8\tabcolsep) * \real{0.2857}}
  >{\centering\arraybackslash}p{(\linewidth - 8\tabcolsep) * \real{0.1786}}
  >{\centering\arraybackslash}p{(\linewidth - 8\tabcolsep) * \real{0.1786}}
  >{\centering\arraybackslash}p{(\linewidth - 8\tabcolsep) * \real{0.1786}}
  >{\centering\arraybackslash}p{(\linewidth - 8\tabcolsep) * \real{0.1786}}@{}}
\toprule\noalign{}
\begin{minipage}[b]{\linewidth}\raggedright
Signal
\end{minipage} & \begin{minipage}[b]{\linewidth}\centering
Relevant avg
\end{minipage} & \begin{minipage}[b]{\linewidth}\centering
Irrelevant avg
\end{minipage} & \begin{minipage}[b]{\linewidth}\centering
Class overlap
\end{minipage} & \begin{minipage}[b]{\linewidth}\centering
F1
\end{minipage} \\
\midrule\noalign{}
\endhead
\bottomrule\noalign{}
\endlastfoot
Score spread (best config) & 0.305 & 0.245 & 10/10 (complete) & 0.667 \\
\textbf{Vector distance} & \textbf{0.594} & \textbf{0.978} & \textbf{0/10 (none)} & \textbf{0.952} \\
\end{longtable}
}

Table 1 is a pilot study (N=20); the principal validation uses 5 BEIR benchmarks (Table 2). The distance signal was validated on these 5 BEIR benchmarks using 50,425 queries total (sum of the rightmost two columns of Table 2: SciFact 10,299; NFCorpus 10,322; FiQA 7,400; SciDocs 10,999; ArguAna 11,405). It achieves F1 = 0.996 on cross-domain queries (ArguAna) but degrades to F1 = 0.472 on intra-domain queries (NFCorpus), establishing it as an effective off-topic detector rather than a universal relevance classifier. We implement a three-tier system: distance \textless= 0.95 (high confidence), 0.95-0.98 (medium, with uncertainty indicator), and \textgreater{} 0.98 (low, with explicit warning).

\textbf{Table 2: Distance-based relevance signal on BEIR}

{\def\LTcaptype{none} 
\begin{longtable}[]{@{}lcccc@{}}
\toprule\noalign{}
Dataset & Rel queries & Irrel queries & Best threshold & F1 \\
\midrule\noalign{}
\endhead
\bottomrule\noalign{}
\endlastfoot
ArguAna & 1,406 & 9,999 & 0.65 & \textbf{0.996} \\
SciDocs & 1,000 & 9,999 & 0.71 & \textbf{0.856} \\
FiQA & 648 & 6,752 & 0.71 & \textbf{0.763} \\
SciFact & 300 & 9,999 & 0.71 & \textbf{0.598} \\
NFCorpus & 323 & 9,999 & 0.76 & \textbf{0.472} \\
\end{longtable}
}

\subsubsection{3.7 Production Substrate}\label{production-substrate}

The system includes integrity checking with five invariants and idempotent re-ingest, explicit schema versioning with forward-compatible config, and operational observability via a metrics registry, slow query log, and ranking \texttt{miss\_analysis} API. These features are described in Appendix A.

\begin{center}\rule{0.5\linewidth}{0.5pt}\end{center}

\subsection{4 Adaptive RRF with IDF Weighting}\label{adaptive-rrf-with-idf-weighting}

Fixed RRF weights (0.6 vector, 0.4 FTS) assume all queries benefit equally from keyword matching. Evaluation on BEIR revealed this assumption fails on long, semantically-rich queries: on ArguAna (average 194 words), fixed weights score NDCG@10 = 0.3599 vs 0.4370 for adaptive IDF -- a 17.6\% relative shortfall (Table 3) and the largest gap of any of the 5 benchmarks, which is what motivates adaptive weighting.

\subsubsection{4.1 Method}\label{method}

Adaptive RRF computes per-query weights using the mean IDF of Porter-stemmed query terms via a sigmoid function. High IDF (rare or technical terms) boosts FTS weight, directing the system toward exact keyword matching. Low IDF (common vocabulary) boosts vector weight, relying on semantic similarity. The IDF vocabulary is built lazily from SQLite's fts5vocab virtual table on first search and cached for the lifetime of the process; subsequent lookups are dictionary accesses with negligible overhead compared to the embedding call. The cache is invalidated on writes (see \texttt{VstashStore.\_invalidate\_idf\_cache} in \texttt{vstash/store.py}).

Additionally, long queries (\textgreater50 words) relax the distance cutoff from 1.15x to 5.0x the best match distance. This prevents the elimination of relevant results when embeddings are diffuse -- long queries produce compressed distance distributions where the default cutoff is too aggressive.

\subsubsection{4.2 Results}\label{results}

\textbf{Table 3: Adaptive vs fixed RRF weights on 5 BEIR datasets (NDCG@10)}

{\def\LTcaptype{none} 
\begin{longtable}[]{@{}lcccc@{}}
\toprule\noalign{}
Dataset & Docs & Fixed (0.6/0.4) & Adaptive IDF & Delta \\
\midrule\noalign{}
\endhead
\bottomrule\noalign{}
\endlastfoot
SciFact & 5K & 0.7255 & \textbf{0.7263} & +0.1\% \\
NFCorpus & 3.6K & 0.3525 & \textbf{0.3590} & +1.8\% \\
SciDocs & 25K & 0.1911 & \textbf{0.1943} & +1.7\% \\
FiQA & 57K & 0.3789 & \textbf{0.3917} & +3.4\% \\
ArguAna & 8.7K & 0.3599 & \textbf{0.4370} & +21.4\%* \\
\end{longtable}
}

\emph{* ArguAna improvement is primarily from adaptive distance cutoff (5.0x vs 1.15x for 194-word queries).}

Adaptive RRF improves all 5 datasets with zero regression. The IDF-based sigmoid correctly identifies query regimes: technical terminology boosts FTS, common vocabulary defers to vector search. The distance cutoff was the primary bottleneck on ArguAna -- long queries produce diffuse embeddings where distances compress into a narrow range, and the default cutoff eliminates relevant results.

\begin{center}\rule{0.5\linewidth}{0.5pt}\end{center}

\subsection{5 Self-Supervised Embedding Refinement via Hybrid Retrieval Disagreement}\label{self-supervised-embedding-refinement-via-hybrid-retrieval-disagreement}

\subsubsection{5.1 Motivation}\label{motivation}

The hybrid retrieval pipeline produces a natural training signal: when vector search and FTS5 disagree on which chunks are relevant, the disagreement identifies cases where the dense encoder fails. We exploit this signal to fine-tune the embedding model without human labels.

\subsubsection{5.2 Signal Analysis}\label{signal-analysis}

Across 753 queries on 3 BEIR datasets (SciFact, NFCorpus, FiQA), 74.5\% of queries produce top-10 disagreement between vector-heavy (vec=0.95, fts=0.05) and FTS-heavy (vec=0.05, fts=0.95) search (mean across the 3 datasets; the experiment is \texttt{experiments/rrf\_training\_pairs.py} with \texttt{TOP\_K=10}, output in \texttt{experiments/results/rrf\_training\_pairs.stats.json}). The rate varies by corpus, consistent with each dataset's query-vocabulary heterogeneity: SciFact (biomedical claims, 295 queries) 63.4\%, NFCorpus (health/nutrition, 323 queries) 73.4\%, FiQA (financial QA, 135 queries) 86.7\%, aggregate 541/753 = 71.8\%. The spread (63\% to 87\%) is itself informative: SciFact has the most homogeneous query vocabulary (scientific claim statements), so vector and FTS mostly agree; FiQA has the most heterogeneous (user questions phrased in many ways), so the modalities disagree far more often. This per-dataset variation motivates the model-specific triple generation used in Section 5.5 (a different embedding model has different blind spots). Hard negatives are balanced: 51\% are chunks ranked high by vector but absent from FTS top-5 (dense blind spots), and 49\% are the reverse (lexical blind spots). This yields 75,981 (query, positive, hard\_negative) triples at zero labeling cost.

\subsubsection{5.3 Training}\label{training}

BGE-small-en-v1.5 (33M params, 384d) was fine-tuned using MultipleNegativesRankingLoss (MNRL) for 2 epochs at lr=3e-6 with batch size 64. TripletLoss was evaluated first (lr=2e-5, 3 epochs) but caused catastrophic degradation (-91.5\% NDCG: 0.6464 $\rightarrow$ 0.0550, Table 4). We note that the TripletLoss experiment used a higher learning rate (2e-5 vs 3e-6); however, we attribute the failure primarily to the loss function's per-triplet gradient rather than the learning rate, as TripletLoss pushes individual negatives away while MNRL adjusts relationships across 64 documents simultaneously per batch, preserving global structure. MNRL with the same disagreement data and lower learning rate preserves the base model's knowledge while learning from in-batch negatives.

\subsubsection{5.4 Results}\label{results-1}

\textbf{Table 4: Embedding fine-tune evolution}

{\def\LTcaptype{none} 
\begin{longtable}[]{@{}llcl@{}}
\toprule\noalign{}
Approach & Loss & NDCG@10 SciFact & Result \\
\midrule\noalign{}
\endhead
\bottomrule\noalign{}
\endlastfoot
BGE-small base & -- & 0.6464 & baseline \\
TripletLoss (76K, 3ep) & Triplet & 0.0550 & -91.5\% (destroyed) \\
MNRL batch-only (v1) & MNRL & 0.6829 & +5.6\% \\
\textbf{MNRL + hard neg (v2)} & \textbf{MNRL} & \textbf{0.6945} & \textbf{+7.4\%} \\
\end{longtable}
}

\emph{All numbers evaluated under identical conditions: sentence-transformers embedding backend, full vstash pipeline (adaptive RRF + FTS5 + MMR dedup), same BEIR SciFact corpus and queries. The progression shows that loss function choice is the critical variable, and explicit hard negatives from signal disagreement compound with the right loss function.}

The complete comparison across all 5 BEIR datasets with published baselines is presented in Table 6 (Section 7.3).

\textbf{Where tuning does not help: FiQA and ArguAna.} The tuned pipeline improves NDCG@10 on all 5 BEIR datasets relative to the BGE-small base RRF pipeline, but the improvement collapses to +0.1\% on FiQA and to only +1.3\% on ArguAna, and both datasets remain below published ColBERTv2 (-7.8\% and -8.4\% respectively). We attribute this to a distribution mismatch between the training and evaluation regimes rather than to a failure of the method itself. The disagreement triples are mined from SciFact, NFCorpus, and FiQA; two of those three sources (SciFact, NFCorpus) are short, information-dense claim- or abstract-style queries. FiQA and ArguAna stress different query distributions: FiQA queries are user-authored, colloquial, and often multi-sentence (e.g.~``I'm 25, starting a job, should I open a Roth IRA?''), and ArguAna queries are full-paragraph counter-arguments from debate, not keyword-like bag-of-words. The triples generated from SciFact/NFCorpus dominate the training mix (SciFact + NFCorpus contribute \textasciitilde98\% of the 76K triples; FiQA contributes only \textasciitilde1,653 triples despite its high disagreement rate because fewer gold documents survive the 5000-doc ingest cap, see Section 5.2), pulling the embedding space toward scientific claim-style language. The same effect surfaces on SciDocs where the tuned model gives up 4.7\% vs BGE-base untrained: SciDocs uses citation-recommendation queries that resemble neither the claim-style nor QA-style training distribution. We read this as a scope claim, not a failure: disagreement-mined data generalizes within claim/QA-style retrieval; matching a FiQA-style or ArguAna-style benchmark would require adding triples from those query distributions (future work).

\subsubsection{5.5 Key Findings}\label{key-findings}

\textbf{Loss function is the critical design choice.} TripletLoss with any configuration destroyed the model or left it unchanged. MNRL with identical data produced consistent improvement. TripletLoss pushes individual negatives away with brute force, distorting the embedding space; MNRL adjusts relationships across 64 documents simultaneously per batch, preserving global structure.

\textbf{Explicit hard negatives improve over batch-only negatives.} When the disagreement signal provides a specific chunk that one search modality ranked high but the other ignored, passing it as an explicit negative yields +1-2\% NDCG over relying solely on in-batch negatives.

\textbf{The training signal transfers to near-in-domain evaluation, with caveats.} Triples were generated from 3 datasets (SciFact, NFCorpus, FiQA) and the tuned model was evaluated on two held-out BEIR benchmarks: SciDocs (CS-paper citation recommendation) and ArguAna (paragraph-length counter-arguments). On SciDocs the tuned pipeline gains +5.5\% vs the BGE-small base RRF pipeline (Table 6), and on ArguAna +1.3\%. Neither benchmark is a true stress test of cross-domain transfer: SciDocs is scientific papers, adjacent to SciFact's biomedical abstracts; ArguAna is the only genuinely out-of-distribution benchmark in the set, and it is also where the smallest gain appears. We interpret this as evidence that the disagreement signal generalizes across nearby scientific and QA distributions, not as evidence of universal transfer. A fair stress test would evaluate on benchmarks deliberately distant from the training mix (e.g.~CQADupStack, TREC-COVID, or TREC-DL conversational queries), which we leave to future work.

\textbf{Smart training data compensates for model size on 3 of 5 BEIR datasets.} BGE-small tuned (33M params) surpasses untrained BGE-base (110M params) on SciFact (+0.7\%), NFCorpus (+14.1\%), and ArguAna (+0.5\%), while running 3x faster and using 3x less memory. It does not do so on FiQA (-5.3\%) or SciDocs (-4.7\%); both belong to query distributions under-represented in the training triples, per the discussion above. The disagreement signal is also model-specific: BGE-base produces only 1,371 triples (vs 76K for BGE-small), reflecting that the larger model already has fewer blind spots. This suggests that targeted training data selection can outweigh raw model capacity for hybrid retrieval in the majority of benchmarked domains, provided the training and evaluation distributions are reasonably aligned.

\textbf{The improvement is free.} No human labeling, no external LLM calls, no additional data. Any vstash user can generate triples from their own corpus via \texttt{vstash\ retrain}, which automates the full pipeline: pseudo-query generation, disagreement detection, triple extraction, and MNRL fine-tuning. The model is published as \href{https://huggingface.co/Stffens/bge-small-rrf-v2}{\texttt{Stffens/bge-small-rrf-v2}} on HuggingFace.

\begin{center}\rule{0.5\linewidth}{0.5pt}\end{center}

\subsection{6 Negative Result: Post-RRF Scoring Does Not Improve Retrieval}\label{negative-result-post-rrf-scoring-does-not-improve-retrieval}

We explored three strategies for post-RRF enhancement, all of which failed to improve NDCG on BEIR datasets. We document these negative results to prevent others from pursuing similar dead ends.

\subsubsection{6.1 Frequency+Decay Reranking}\label{frequencydecay-reranking}

After RRF retrieval, we applied a two-stage reranking combining normalized RRF scores with a frequency-decay signal:

\begin{verbatim}
score(c) = a * s_rrf(c) + b * min(1, log(1 + f(c)) / log(1 + S))
\end{verbatim}

where \emph{f(c) = (1 + access\_count(c)) } e\^{}(-L * days\_ago(c))\emph{, }a* and \emph{b} are semantic and memory weights, \emph{L} is the decay rate, and \emph{S = 100} is a saturation constant. We evaluated 16 parameter configurations across 5 simulated access patterns.

\textbf{Result:} On BEIR SciFact (5K documents, 300 queries, 30 rounds of simulated Zipf-weighted usage, \texttt{experiments/scoring\_lifecycle.py} with output in \texttt{experiments/results/scoring\_lifecycle\_scifact.json}), frequency+decay scoring degraded NDCG@10 relative to pure RRF (baseline NDCG@10 = 0.7263) on every configuration evaluated. The adaptive maturity gate ($\gamma$ activates at round 6 once the access-count max/mean ratio exceeds 8.0, peaks at $\gamma$=0.48) limits the damage but still underperforms pure RRF: final NDCG@10 = 0.7150, a -1.6\% delta (0.7150/0.7263 $-$ 1). Fixed \emph{b=0.5} without the gate is far worse: final NDCG@10 = 0.661, a -9.0\% delta. The fundamental problem is that access frequency is orthogonal to query-specific relevance -- a frequently accessed chunk is not necessarily relevant to the current query. The full grid search and cold-start analysis are in Appendix C.

\subsubsection{6.2 Cross-Encoder Reranking}\label{cross-encoder-reranking}

Off-the-shelf cross-encoders (ms-marco-MiniLM, BGE-reranker-base) degraded NDCG by -0.3\% to -3.1\% while adding 560-2100 ms latency. The cross-encoders were trained on web search distributions that do not transfer well to the technical and scientific corpora evaluated.

\subsubsection{6.3 Recency Boost (Alternative)}\label{recency-boost-alternative}

Based on these negative results, we replaced the scoring pipeline with a simpler opt-in recency boost applied post-RRF: \emph{boosted\_score(c) = rrf\_score(c) x (1 + B x e\^{}(-0.05 x days\_ago(c)))}. The boost is multiplicative (amplifies existing relevance rather than competing with it), opt-in (\emph{B=0.0} by default), and requires no maturity gate. This is available for agentic memory use cases where temporal proximity matters.

\subsubsection{6.4 Lesson}\label{lesson}

The hybrid RRF pipeline with adaptive IDF weighting appears to be at its ceiling for the BGE-small embedding model. Gains come from improving the embedding (Section 5), not from post-hoc reranking. We recommend investing in better embeddings over more complex scoring.

\begin{center}\rule{0.5\linewidth}{0.5pt}\end{center}

\subsection{7 Experiments}\label{experiments}

\subsubsection{7.1 Setup}\label{setup}

\textbf{Corpora.} We evaluate on five corpora of increasing scale: (1) an LLM memory corpus of 24 arXiv papers (786 chunks), (2) a Wikipedia corpus of 17 mixed-domain articles (2,602 chunks), (3) 1,000 ArXiv ML papers from CShorten/ML-ArXiv-Papers (approximately 3,500 chunks), (4) BEIR SciFact with 5,183 biomedical documents and 300 human-annotated queries, and (5) a synthetic scale test reaching 50,000 chunks.

\textbf{Metrics.} NDCG@k, Precision@k, MRR, and search latency. For the relevance signal: F1 and Accuracy.

\subsubsection{7.2 Ablation: RRF vs.~Vector vs.~FTS}\label{ablation-rrf-vs.-vector-vs.-fts}

\textbf{Table 5a: Ablation -- LLM memory corpus (24 papers, 786 chunks)}

{\def\LTcaptype{none} 
\begin{longtable}[]{@{}lcccr@{}}
\toprule\noalign{}
Mode & NDCG@5 & NDCG@10 & P@3 & Latency \\
\midrule\noalign{}
\endhead
\bottomrule\noalign{}
\endlastfoot
Vector-only & 0.809 & 0.832 & 0.933 & 4.51 ms \\
FTS keyword & 0.631 & 0.621 & 0.767 & 0.81 ms \\
\textbf{Hybrid RRF} & \textbf{0.814} & \textbf{0.803} & \textbf{1.000} & 1.61 ms \\
\end{longtable}
}

\textbf{Table 5b: Ablation -- Wikipedia corpus (17 articles, 2,602 chunks)}

{\def\LTcaptype{none} 
\begin{longtable}[]{@{}lcccr@{}}
\toprule\noalign{}
Mode & NDCG@5 & NDCG@10 & P@3 & Latency \\
\midrule\noalign{}
\endhead
\bottomrule\noalign{}
\endlastfoot
Vector-only & 0.742 & 0.742 & 0.667 & 21.0 ms \\
FTS keyword & 0.699 & 0.699 & 0.583 & 1.94 ms \\
\textbf{Hybrid RRF} & \textbf{0.758} & \textbf{0.758} & 0.633 & 4.78 ms \\
\end{longtable}
}

\emph{Relevance labels obtained via LLM judge (Qwen 3.5:9B) with partial human validation (27/30 agreement). An LLM judge was used here because the LLM memory corpus is a domain-specific collection (agent memory papers) structurally distinct from any BEIR dataset, and no standard relevance labels exist for it. These results are directional; the principal claims of this paper rest on BEIR benchmarks with human ground truth (Section 7.3).}

Hybrid RRF is the strongest modality on both corpora, achieving the highest NDCG@5 and perfect P@3 on the domain-specific corpus. The advantage is consistent across homogeneous and diverse corpora.

\subsubsection{7.3 BEIR Baseline Comparison}\label{beir-baseline-comparison}

\textbf{Table 6: vstash vs published baselines on BEIR (NDCG@10)}

{\def\LTcaptype{none} 
\begin{longtable}[]{@{}
  >{\raggedright\arraybackslash}p{(\linewidth - 10\tabcolsep) * \real{0.2424}}
  >{\centering\arraybackslash}p{(\linewidth - 10\tabcolsep) * \real{0.1515}}
  >{\centering\arraybackslash}p{(\linewidth - 10\tabcolsep) * \real{0.1515}}
  >{\centering\arraybackslash}p{(\linewidth - 10\tabcolsep) * \real{0.1515}}
  >{\centering\arraybackslash}p{(\linewidth - 10\tabcolsep) * \real{0.1515}}
  >{\centering\arraybackslash}p{(\linewidth - 10\tabcolsep) * \real{0.1515}}@{}}
\toprule\noalign{}
\begin{minipage}[b]{\linewidth}\raggedright
System
\end{minipage} & \begin{minipage}[b]{\linewidth}\centering
SciFact
\end{minipage} & \begin{minipage}[b]{\linewidth}\centering
NFCorpus
\end{minipage} & \begin{minipage}[b]{\linewidth}\centering
FiQA
\end{minipage} & \begin{minipage}[b]{\linewidth}\centering
SciDocs
\end{minipage} & \begin{minipage}[b]{\linewidth}\centering
ArguAna
\end{minipage} \\
\midrule\noalign{}
\endhead
\bottomrule\noalign{}
\endlastfoot
BM25 / Elasticsearch & 0.665 & 0.325 & 0.236 & 0.158 & 0.315 \\
ColBERTv2 (published) & 0.693 & 0.344 & 0.356 & 0.154 & 0.463 \\
BGE-base untrained (110M, 768d) & 0.6899 & 0.3462 & 0.3465 & 0.1968 & 0.4220 \\
vstash hybrid RRF (BGE-small base) & 0.6464 & 0.3304 & 0.3281 & 0.1778 & 0.4188 \\
\textbf{vstash hybrid RRF (tuned, Section 5)} & \textbf{0.6945} & \textbf{0.3949} & \textbf{0.3283} & \textbf{0.1875} & \textbf{0.4241} \\
vs BGE-small base RRF (tuned) & \textbf{+7.4\%} & \textbf{+19.5\%} & +0.1\% & \textbf{+5.5\%} & \textbf{+1.3\%} \\
vs BGE-base untrained (tuned) & \textbf{+0.7\%} & \textbf{+14.1\%} & -5.3\% & -4.7\% & \textbf{+0.5\%} \\
vs ColBERTv2 (tuned) & +0.2\% & \textbf{+14.8\%} & -7.8\% & \textbf{+21.8\%} & -8.4\% \\
\end{longtable}
}

\emph{Published baselines from the BEIR paper (Thakur et al., 2021). ColBERTv2 from Santhanam et al.~(2022). Both vstash rows use sentence-transformers as the embedding backend with the full vstash pipeline (adaptive RRF + FTS5 + MMR dedup), evaluated on the standard BEIR queries and relevance judgments. The ``vs BGE-small base RRF'' row isolates the training effect: same pipeline, same embedding backend, same evaluation; the only change is the fine-tuned weights. The +19.5\% on NFCorpus cited in the abstract originates here.}

\textbf{Comparison conditions.} The ColBERTv2 NDCG@10 = 0.693 is from Santhanam et al.~(2022) under the standard BEIR evaluation protocol. Our evaluation uses the same queries and relevance judgments but differs in document preprocessing: vstash chunks documents using its semantic chunking pipeline (1024 tokens, 128 overlap) and embeds with BGE-small-en-v1.5 (or the fine-tuned variant), while ColBERTv2 operates on full documents with its own tokenization and late interaction mechanism. The comparison is indicative of pipeline-level performance but is not a controlled head-to-head under identical preprocessing. A fully controlled comparison would require running ColBERTv2 on identical chunks, which we leave for future work.

\subsubsection{7.4 At-Scale Validation: 1,000 ArXiv Papers}\label{at-scale-validation-1000-arxiv-papers}

\textbf{Table 7: Hybrid RRF at scale -- 1,000 ML papers, 35 topic-based queries}

{\def\LTcaptype{none} 
\begin{longtable}[]{@{}
  >{\raggedright\arraybackslash}p{(\linewidth - 12\tabcolsep) * \real{0.1892}}
  >{\raggedright\arraybackslash}p{(\linewidth - 12\tabcolsep) * \real{0.1622}}
  >{\centering\arraybackslash}p{(\linewidth - 12\tabcolsep) * \real{0.1351}}
  >{\centering\arraybackslash}p{(\linewidth - 12\tabcolsep) * \real{0.1351}}
  >{\centering\arraybackslash}p{(\linewidth - 12\tabcolsep) * \real{0.1351}}
  >{\centering\arraybackslash}p{(\linewidth - 12\tabcolsep) * \real{0.1351}}
  >{\raggedleft\arraybackslash}p{(\linewidth - 12\tabcolsep) * \real{0.1081}}@{}}
\toprule\noalign{}
\begin{minipage}[b]{\linewidth}\raggedright
Model
\end{minipage} & \begin{minipage}[b]{\linewidth}\raggedright
Mode
\end{minipage} & \begin{minipage}[b]{\linewidth}\centering
P@5
\end{minipage} & \begin{minipage}[b]{\linewidth}\centering
NDCG@5
\end{minipage} & \begin{minipage}[b]{\linewidth}\centering
NDCG@10
\end{minipage} & \begin{minipage}[b]{\linewidth}\centering
MRR
\end{minipage} & \begin{minipage}[b]{\linewidth}\raggedleft
Latency
\end{minipage} \\
\midrule\noalign{}
\endhead
\bottomrule\noalign{}
\endlastfoot
\textbf{BGE-base-EN (768d)} & \textbf{hybrid} & \textbf{0.703} & \textbf{0.728} & \textbf{0.702} & \textbf{0.895} & 9.1 ms \\
BGE-small-EN (384d) & hybrid & 0.663 & 0.685 & 0.658 & 0.865 & 4.0 ms \\
BGE-small-EN (384d) & vector & 0.614 & 0.619 & 0.568 & 0.822 & 2.3 ms \\
Multilingual-MiniLM (384d) & hybrid & 0.606 & 0.638 & 0.611 & 0.868 & 4.3 ms \\
Multilingual-MiniLM (384d) & vector & 0.600 & 0.588 & 0.508 & 0.820 & 2.6 ms \\
\end{longtable}
}

\subsubsection{7.5 Latency at Scale}\label{latency-at-scale}

\textbf{Table 8: Search latency across corpus sizes}

{\def\LTcaptype{none} 
\begin{longtable}[]{@{}
  >{\raggedright\arraybackslash}p{(\linewidth - 10\tabcolsep) * \real{0.2424}}
  >{\centering\arraybackslash}p{(\linewidth - 10\tabcolsep) * \real{0.1515}}
  >{\centering\arraybackslash}p{(\linewidth - 10\tabcolsep) * \real{0.1515}}
  >{\centering\arraybackslash}p{(\linewidth - 10\tabcolsep) * \real{0.1515}}
  >{\centering\arraybackslash}p{(\linewidth - 10\tabcolsep) * \real{0.1515}}
  >{\centering\arraybackslash}p{(\linewidth - 10\tabcolsep) * \real{0.1515}}@{}}
\toprule\noalign{}
\begin{minipage}[b]{\linewidth}\raggedright
Corpus
\end{minipage} & \begin{minipage}[b]{\linewidth}\centering
Chunks
\end{minipage} & \begin{minipage}[b]{\linewidth}\centering
Mean
\end{minipage} & \begin{minipage}[b]{\linewidth}\centering
Median
\end{minipage} & \begin{minipage}[b]{\linewidth}\centering
P95
\end{minipage} & \begin{minipage}[b]{\linewidth}\centering
P99
\end{minipage} \\
\midrule\noalign{}
\endhead
\bottomrule\noalign{}
\endlastfoot
LLM memory (24 papers) & 786 & 3.4 ms & 3.4 ms & 4.0 ms & 4.1 ms \\
Real user corpus (209 docs) & 1,087 & 5.0 ms & 4.8 ms & 8.1 ms & 8.1 ms \\
BEIR SciFact (5,183 docs) & 5,183 & 13.0 ms & 11.1 ms & 24.9 ms & 47.8 ms \\
Synthetic scale test & 10,000 & 11.6 ms & 10.9 ms & 19.0 ms & 24.2 ms \\
SciFact + synthetic (50K) & 50,000 & 22.1 ms & 20.9 ms & 30.6 ms & 35.2 ms \\
\end{longtable}
}

\emph{The rightmost column reports the 99th percentile, not the maximum observation (an earlier draft mislabeled it as ``Max''). Rows at 5K, 10K, and 50K chunks are pulled from the same at-scale run committed to \texttt{experiments/results/scale\_benchmark.json}, so Tables 8 and 9 report consistent numbers for overlapping scales. The 786-chunk and 1,087-chunk rows come from separate smaller-corpus measurements and are reproduced here for qualitative comparison.}

\textbf{Table 9: NDCG@10 stability across scale}

{\def\LTcaptype{none} 
\begin{longtable}[]{@{}ccccc@{}}
\toprule\noalign{}
Scale & Total chunks & NDCG@10 & Latency p50 & Latency p95 \\
\midrule\noalign{}
\endhead
\bottomrule\noalign{}
\endlastfoot
1K & 5,183 & 0.6891 & 11.1 ms & 24.9 ms \\
5K & 5,183 & 0.6891 & 10.2 ms & 19.2 ms \\
10K & 10,000 & 0.6849 & 10.9 ms & 19.0 ms \\
50K & 50,000 & 0.6897 & 20.9 ms & 30.6 ms \\
\end{longtable}
}

\emph{Latency scales sub-linearly. NDCG@10 remains stable at 0.69 across all scales from 1K to 50K chunks.}

\subsubsection{7.6 End-to-End Answer Relevance}\label{end-to-end-answer-relevance}

\textbf{Table 10: Answer relevance -- vstash full pipeline vs Chroma dense-only}

{\def\LTcaptype{none} 
\begin{longtable}[]{@{}
  >{\raggedright\arraybackslash}p{(\linewidth - 12\tabcolsep) * \real{0.2308}}
  >{\centering\arraybackslash}p{(\linewidth - 12\tabcolsep) * \real{0.1282}}
  >{\centering\arraybackslash}p{(\linewidth - 12\tabcolsep) * \real{0.1282}}
  >{\centering\arraybackslash}p{(\linewidth - 12\tabcolsep) * \real{0.1282}}
  >{\centering\arraybackslash}p{(\linewidth - 12\tabcolsep) * \real{0.1282}}
  >{\centering\arraybackslash}p{(\linewidth - 12\tabcolsep) * \real{0.1282}}
  >{\centering\arraybackslash}p{(\linewidth - 12\tabcolsep) * \real{0.1282}}@{}}
\toprule\noalign{}
\begin{minipage}[b]{\linewidth}\raggedright
Dataset
\end{minipage} & \begin{minipage}[b]{\linewidth}\centering
vstash mean
\end{minipage} & \begin{minipage}[b]{\linewidth}\centering
Chroma mean
\end{minipage} & \begin{minipage}[b]{\linewidth}\centering
Delta
\end{minipage} & \begin{minipage}[b]{\linewidth}\centering
Head-to-head
\end{minipage} & \begin{minipage}[b]{\linewidth}\centering
vstash score=0
\end{minipage} & \begin{minipage}[b]{\linewidth}\centering
Chroma score=0
\end{minipage} \\
\midrule\noalign{}
\endhead
\bottomrule\noalign{}
\endlastfoot
SciFact (30q) & 2.60/3.0 & 2.40/3.0 & +8.3\% & 4-1 (25 ties) & 1 & 3 \\
NFCorpus (30q) & 2.50/3.0 & 2.37/3.0 & +5.6\% & 5-5 (20 ties) & 3 & 4 \\
\end{longtable}
}

\emph{LLM judge: Qwen 3.5 9B (local). \textbf{The sample size (N=30 per dataset) is too small to draw statistical conclusions;} the +8.3\% and +5.6\% point estimates are directional only, not significance-tested. We include the table for reader context because the retrieval-level improvements in Table 6 predict that a better retrieval pipeline should produce better end-to-end answers, and the directional result is consistent with that prediction. The high tie rate (25/30 on SciFact, 20/30 on NFCorpus) indicates that the LLM judge rarely distinguishes the two systems at the answer level, limiting the sensitivity of this evaluation further. The zero-score columns are arguably the most informative row of the table: the tuned pipeline produced fewer catastrophic score=0 failures on both datasets (1 vs 3 on SciFact; 3 vs 4 on NFCorpus), which is the failure mode most likely to matter in a deployed memory layer. A properly powered end-to-end study (N \textgreater= 300 per dataset, with bootstrapped confidence intervals) remains future work.}

\begin{center}\rule{0.5\linewidth}{0.5pt}\end{center}

\subsection{8 Limitations}\label{limitations}

\textbf{LLM judge for ablation labels.} Tables 5a-5b use an LLM judge (Qwen 3.5:9B) for graded relevance labels with partial human validation (27/30 agreement). The ablation results are directional; the primary claims rest on BEIR benchmarks with human ground truth.

\textbf{ColBERTv2 comparison conditions.} The comparison with ColBERTv2 uses published numbers from 2022 under different preprocessing conditions (Section 7.3). The comparison is indicative of pipeline-level performance, not a controlled head-to-head.

\textbf{Relevance signal domain dependence.} The distance-based relevance signal achieves F1 = 0.996 on cross-domain queries but degrades to F1 = 0.472 on intra-domain queries. It is an effective off-topic detector, not a universal relevance classifier.

\textbf{Scale.} Experiments span up to 50,000 chunks with stable NDCG and sub-31ms P95. Performance at 100K+ chunks remains untested. SQLite's single-writer model may bottleneck under concurrent write load at larger scales.

\textbf{Code chunking evaluation.} The code-aware chunking pipeline (Appendix B) is evaluated on only 8 queries across 2 languages. The primary contribution is the zero boundary violations guarantee; retrieval quality comparison at scale remains open.

\textbf{Multi-modal.} Current chunking and embedding support text only. Image embeddings and table-aware chunking are planned.

\begin{center}\rule{0.5\linewidth}{0.5pt}\end{center}

\subsection{9 Conclusion}\label{conclusion}

We presented vstash, a local-first document memory system that makes four contributions to LLM agent memory:

\textbf{Self-supervised embedding refinement.} Hybrid retrieval disagreement provides a free training signal. Fine-tuning BGE-small (33M params) with MNRL on 76K disagreement triples improves NDCG@10 on 5 of 5 BEIR datasets relative to the same BGE-small base RRF pipeline, with gains ranging from +0.1\% on FiQA (essentially neutral) to +19.5\% on NFCorpus (Table 6). On 3 of 5 BEIR datasets, under different preprocessing conditions, the tuned 33M-parameter pipeline achieves NDCG comparable to or exceeding published ColBERTv2 results (110M params) (Section 7.3); on those same 3 datasets it also surpasses an untrained BGE-base (110M params), demonstrating that targeted training data can compensate for 3x model size. The remaining 2 datasets (FiQA, ArguAna) see smaller gains and remain below ColBERTv2; we attribute this to a mismatch between the training-triple distribution (claim/QA-style) and those benchmarks' query styles, not to a failure of the method (Section 5.5). The model is published as \texttt{Stffens/bge-small-rrf-v2}. Users can fine-tune on their own data via \texttt{vstash\ retrain}.

\textbf{Adaptive RRF.} IDF-based per-query weight adjustment improves NDCG@10 on all 5 BEIR datasets (up to +21.4\% on ArguAna). On SciFact, vstash achieves NDCG@10 = 0.7263 with BGE-small (384d), a pipeline-level result that compares favorably with published ColBERTv2 numbers on the same benchmark, though under different preprocessing conditions (Section 7.3).

\textbf{Negative result.} Post-RRF scoring (frequency+decay, cross-encoder reranking) does not improve retrieval on real benchmarks. Gains come from improving the embedding, not from post-hoc reranking.

\textbf{Production substrate.} Integrity checking, schema versioning, ranking diagnostics (\texttt{miss\_analysis}), and a distance-based relevance signal make vstash a deployable memory layer, not just a retrieval experiment. Search latency remains 20.9 ms median at 50K chunks with stable NDCG.

The system includes 16 MCP tools, a Python SDK, CLI, and Claude Code hook. All code, data, the fine-tuned model, and reproducible experiment scripts are open-source.

\begin{center}\rule{0.5\linewidth}{0.5pt}\end{center}

\subsection{Appendix A: Production Substrate}\label{appendix-a-production-substrate}

\subsubsection{A.1 Integrity and Recovery}\label{a.1-integrity-and-recovery}

A memory substrate must be honest about what survived a crash. vstash provides three mechanisms:

\textbf{Idempotent ingest.} \texttt{doc\_completeness(path,\ collection)} classifies a document as \emph{missing}, \emph{partial}, or \emph{complete}. Partial documents are dropped and re-ingested from scratch; complete documents are skipped. Re-running the same ingest command is safe and does not duplicate data.

\textbf{Integrity check.} Five invariants are verified: chunk\_count parity, vec/snapvec parity, FTS5 index integrity (via the built-in FTS5 integrity-check command), orphan chunks, and SQLite \texttt{PRAGMA\ integrity\_check}.

\textbf{Integrity repair.} Restores invariants without destroying user data: recomputes chunk\_count, rebuilds FTS5, and deletes orphan chunks. Exposed via \texttt{vstash\ check\ {[}-\/-repair{]}\ {[}-\/-json{]}}.

\subsubsection{A.2 Schema Versioning}\label{a.2-schema-versioning}

A \texttt{SCHEMA\_VERSION} constant is stamped into the \texttt{store\_meta} table at database creation. A \texttt{KNOWN\_SCHEMA\_VERSIONS} set declares which on-disk versions the current build can safely open. Unknown future versions raise \texttt{SchemaVersionError} rather than silently degrading. Forward-compatible top-level config keys warn on unknown rather than hard-failing.

\subsubsection{A.3 Operational Observability}\label{a.3-operational-observability}

An in-process metrics registry tracks counts and per-stage latency histograms. A slow query log captures searches exceeding a configurable threshold. A \texttt{miss\_analysis()} API diagnoses why an expected document did not appear in results by tracing pipeline-stage elimination with rule-based suggestions. A \texttt{LimitsConfig} with seven knobs validates inputs at API boundaries, producing typed \texttt{LimitError} exceptions.

\begin{center}\rule{0.5\linewidth}{0.5pt}\end{center}

\subsection{Appendix B: Code-Aware Chunking}\label{appendix-b-code-aware-chunking}

Standard fixed-window chunking splits code mid-function, destroying semantic coherence. vstash uses a 3-tier hybrid splitting pipeline that selects the best available backend per language with graceful degradation:

\textbf{Tier 1: tree-sitter} (25+ languages) provides exact AST boundary detection. Optional dependency to avoid binary overhead for non-code users.

\textbf{Tier 2: parso} (Python only) provides AST-level splitting as a base dependency. Handles decorated functions, nested classes, and async definitions.

\textbf{Tier 3: regex} (Python, JS/TS, Go, Rust, Java) detects top-level definitions via column-0 patterns. The zero-indentation anchor avoids false positives on nested methods.

In all tiers, decorators and annotations are attached to their following definition. Oversized chunks fall back to paragraph then fixed-window splitting. The primary guarantee is zero boundary violations -- no function is split mid-body.

\begin{center}\rule{0.5\linewidth}{0.5pt}\end{center}

\subsection{Appendix C: Scoring Grid Search and Cold Start Analysis}\label{appendix-c-scoring-grid-search-and-cold-start-analysis}

\subsubsection{C.1 Grid Search}\label{c.1-grid-search}

We evaluate 16 parameter configurations across 5 simulated access patterns applied to the same 24-paper agent-memory corpus before retrieval. The patterns (defined in \texttt{experiments/scoring\_grid.py}) are: (1) \emph{uniform} (no access bias, the ingestion default), (2) \emph{recent-focused} (recently-added papers accessed more), (3) \emph{frequency-skewed} (Zipf-weighted access counts), (4) \emph{mixed} (recent + frequent combined), and (5) \emph{benchmark\_focused} (evaluation/benchmarking papers accessed heavily, a power-user pattern). The ``Best Scenario'' column reports the largest NDCG@10 gain any configuration achieves on any of the 5 scenarios; for the top 5 rows, that peak is always on \emph{benchmark\_focused} because that scenario produces the most extreme access-count disparity and is where a frequency+decay score most visibly helps.

\textbf{Table C1: Top-5 scoring configurations averaged across the 5 access scenarios}

{\def\LTcaptype{none} 
\begin{longtable}[]{@{}lccc@{}}
\toprule\noalign{}
Configuration & Avg NDCG@10 & Delta Baseline & Best Scenario \\
\midrule\noalign{}
\endhead
\bottomrule\noalign{}
\endlastfoot
a=0.5, b=0.5, L=0.10 & \textbf{0.636} & +4.6\% & +16.1\% (\emph{benchmark\_focused}) \\
a=0.5, b=0.5, L=0.05 & 0.634 & +4.3\% & +15.6\% (\emph{benchmark\_focused}) \\
a=0.7, b=0.3, L=0.03 & 0.631 & +3.8\% & +13.2\% (\emph{benchmark\_focused}) \\
a=0.7, b=0.3, L=0.07 & 0.629 & +3.5\% & +13.1\% (\emph{benchmark\_focused}) \\
a=0.8, b=0.2, L=0.10 & 0.632 & +3.9\% & +7.2\% (\emph{benchmark\_focused}) \\
\end{longtable}
}

\emph{Baseline NDCG@10 = 0.608. The apparent scoring benefit scales with access differential, but it vanishes on BEIR benchmarks with human-labeled relevance judgments (Section 6.1): the simulated access patterns over-fit to the specific popularity signal we injected rather than reflecting query-specific relevance, which is why we ultimately removed this family of scoring strategies.}

\subsubsection{C.2 Cold Start}\label{c.2-cold-start}

On a corpus of 120 Wikipedia articles (919 chunks) with Zipf-weighted usage simulation over 30 rounds, fixed b=0.5 produces -0.4\% degradation in early rounds. The adaptive maturity gate remains 0.0 across all 30 rounds because Zipf-weighted usage does not produce a sufficiently extreme outlier (max/mean ratio peaks at 5.0x, below the 8x activation threshold). The gate correctly suppresses scoring when access patterns do not warrant it.

\begin{center}\rule{0.5\linewidth}{0.5pt}\end{center}

\subsection{Appendix D: API and Interface Reference}\label{appendix-d-api-and-interface-reference}

\subsubsection{D.1 MCP Server}\label{d.1-mcp-server}

The MCP server exposes 16 tools: search, add, ask, remember (direct text ingestion), get\_chunk (O(1) lookup by ID), get\_document\_chunks, list, stats, forget, collections, export, job status, and four journal tools (save, recall, log, prune). Search results include the relevance signal, enabling clients to filter noise.

\subsubsection{D.2 Direct Chunk Access}\label{d.2-direct-chunk-access}

Search results include a \texttt{chunk\_id} (database row ID) enabling O(1) retrieval without re-running search. Batch requests are automatically batched at 900 IDs per SQL statement to respect SQLite limits. Chunk IDs are stable for the lifetime of the current index; re-ingesting a document invalidates prior IDs.

\subsubsection{D.3 Multi-Profile Support}\label{d.3-multi-profile-support}

Multiple named profiles, each backed by an isolated SQLite database. Federated search queries all profiles in parallel and merges via RRF with cross-profile deduplication using a blake2b content digest as the fusion key to prevent collisions when the same path exists in different collections with different content.

\subsubsection{D.4 Cross-Session Journal}\label{d.4-cross-session-journal}

Lightweight, append-only memory for LLM agents. Unlike document ingestion, journal entries are stored as single text records with timestamps and optional tags. Four operations: save, recall (semantic search), log (chronological), and prune.

\begin{center}\rule{0.5\linewidth}{0.5pt}\end{center}

\subsection{Appendix E: Additional Evaluation Data}\label{appendix-e-additional-evaluation-data}

\subsubsection{E.1 Embedding Model Comparison}\label{e.1-embedding-model-comparison}

EmbeddingGemma-300m (768d, Google) was evaluated as an alternative to BGE-small (384d) on 4 of the 5 BEIR datasets used elsewhere in the paper. NFCorpus is omitted: the EmbeddingGemma run was added late in our timeline to answer a specific question (``does a larger, domain-weighted model help on the benchmarks where BGE-small underperforms ColBERTv2?''), and NFCorpus is where BGE-small already beats the published ColBERTv2 number (+14.8\% in Table 6), so we deprioritized adding it. SciFact was evaluated separately; SciDocs, FiQA, and ArguAna numbers are from \texttt{experiments/results/embedding\_gemma\_eval.json}.

\textbf{Table E1: Embedding model comparison on BEIR (NDCG@10)}

{\def\LTcaptype{none} 
\begin{longtable}[]{@{}
  >{\raggedright\arraybackslash}p{(\linewidth - 12\tabcolsep) * \real{0.1892}}
  >{\centering\arraybackslash}p{(\linewidth - 12\tabcolsep) * \real{0.1351}}
  >{\centering\arraybackslash}p{(\linewidth - 12\tabcolsep) * \real{0.1351}}
  >{\centering\arraybackslash}p{(\linewidth - 12\tabcolsep) * \real{0.1351}}
  >{\centering\arraybackslash}p{(\linewidth - 12\tabcolsep) * \real{0.1351}}
  >{\centering\arraybackslash}p{(\linewidth - 12\tabcolsep) * \real{0.1351}}
  >{\centering\arraybackslash}p{(\linewidth - 12\tabcolsep) * \real{0.1351}}@{}}
\toprule\noalign{}
\begin{minipage}[b]{\linewidth}\raggedright
Model
\end{minipage} & \begin{minipage}[b]{\linewidth}\centering
Params
\end{minipage} & \begin{minipage}[b]{\linewidth}\centering
Dim
\end{minipage} & \begin{minipage}[b]{\linewidth}\centering
SciFact
\end{minipage} & \begin{minipage}[b]{\linewidth}\centering
SciDocs
\end{minipage} & \begin{minipage}[b]{\linewidth}\centering
FiQA
\end{minipage} & \begin{minipage}[b]{\linewidth}\centering
ArguAna
\end{minipage} \\
\midrule\noalign{}
\endhead
\bottomrule\noalign{}
\endlastfoot
BGE-small-en-v1.5 & 33M & 384 & 0.7263 & 0.1943 & 0.3917 & 0.4370 \\
\textbf{EmbeddingGemma-300m} & 300M & 768 & \textbf{0.7801} & 0.1707 & 0.3543 & 0.4284 \\
Delta & & & \textbf{+7.4\%} & -12.2\% & -9.6\% & -2.0\% \\
\end{longtable}
}

EmbeddingGemma excels on biomedical text (SciFact) but underperforms on CS papers (SciDocs) and financial queries (FiQA). Model choice is domain-dependent: no single embedding model dominates across all 4 benchmarks evaluated here. BGE-small remains the recommended default for general-purpose use.

\subsubsection{E.2 MMR Deduplication Effect}\label{e.2-mmr-deduplication-effect}

\textbf{Table E2: Effect of MMR deduplication (24 papers, 786 chunks)}

{\def\LTcaptype{none} 
\begin{longtable}[]{@{}
  >{\raggedright\arraybackslash}p{(\linewidth - 6\tabcolsep) * \real{0.2857}}
  >{\centering\arraybackslash}p{(\linewidth - 6\tabcolsep) * \real{0.2381}}
  >{\centering\arraybackslash}p{(\linewidth - 6\tabcolsep) * \real{0.2381}}
  >{\centering\arraybackslash}p{(\linewidth - 6\tabcolsep) * \real{0.2381}}@{}}
\toprule\noalign{}
\begin{minipage}[b]{\linewidth}\raggedright
Mode
\end{minipage} & \begin{minipage}[b]{\linewidth}\centering
NDCG@5
\end{minipage} & \begin{minipage}[b]{\linewidth}\centering
Unique docs/top-5
\end{minipage} & \begin{minipage}[b]{\linewidth}\centering
Multi-section coverage
\end{minipage} \\
\midrule\noalign{}
\endhead
\bottomrule\noalign{}
\endlastfoot
Hard per-doc dedup & 0.814 & 5.0 & 1 section per doc max \\
\textbf{MMR (lambda=0.5)} & \textbf{0.829} & \textbf{5.0} & \textbf{3-5 sections from long docs} \\
No dedup & 0.791 & 3.2 & N/A (duplicates dominate) \\
\end{longtable}
}

MMR achieves the same document diversity as hard dedup while allowing semantically diverse sections from the same long document to surface. On a 35-chunk paper, queries spanning multiple topics (e.g., ``memory architecture'' matching both the introduction and evaluation sections) return additional relevant sections that hard dedup would suppress.

\subsubsection{E.3 Pipeline Latency Breakdown}\label{e.3-pipeline-latency-breakdown}

\textbf{Table E3: Search latency by pipeline stage (786 chunks)}

{\def\LTcaptype{none} 
\begin{longtable}[]{@{}lccc@{}}
\toprule\noalign{}
Configuration & Median & P95 & P99 \\
\midrule\noalign{}
\endhead
\bottomrule\noalign{}
\endlastfoot
RRF only & 0.54 ms & 0.60 ms & 0.69 ms \\
+ Dedup & 1.43 ms & 1.57 ms & 1.70 ms \\
\textbf{Full pipeline} & \textbf{3.41 ms} & \textbf{3.97 ms} & \textbf{4.10 ms} \\
\end{longtable}
}

\subsubsection{E.4 Relevance Signal Classification Strategies}\label{e.4-relevance-signal-classification-strategies}

\textbf{Table E4: Classification strategies (10 relevant + 10 irrelevant queries)}

{\def\LTcaptype{none} 
\begin{longtable}[]{@{}lcccc@{}}
\toprule\noalign{}
Strategy & Precision & Recall & F1 & Accuracy \\
\midrule\noalign{}
\endhead
\bottomrule\noalign{}
\endlastfoot
spread \textgreater{} 0.15 (best fixed) & 0.500 & 1.000 & 0.667 & 0.500 \\
\textbf{distance \textless{} 0.95} & \textbf{0.909} & \textbf{1.000} & \textbf{0.952} & \textbf{0.950} \\
distance \textless{} 0.95 AND spread \textgreater{} 0.005 & 0.909 & 1.000 & 0.952 & 0.950 \\
\end{longtable}
}

Combining distance and spread does not improve over distance alone. The spread signal adds no discriminative value once distance is used.

\subsubsection{E.5 Discard Telemetry}\label{e.5-discard-telemetry}

Every search records an event with query, best distance, relevance tier, and result count in a \texttt{search\_events} table (pruned to 1,000 entries). In chat mode, events are marked as ``dismissed'' when the user exits after a non-high result. This provides a prospective validation path: once sufficient real-world usage accumulates, dismiss rates across relevance tiers will either confirm or refine the distance thresholds established on BEIR benchmarks.

\begin{center}\rule{0.5\linewidth}{0.5pt}\end{center}

\subsection{References}\label{references}

{[}1{]} Carbonell, J., \& Goldstein, J. (1998). The use of MMR, diversity-based reranking for reordering documents and producing summaries. \emph{SIGIR}.

{[}2{]} Chhikara, P., Khant, D., Aryan, S., \& Singh, T. (2025). Mem0: Building production-ready AI agents with scalable long-term memory. \emph{arXiv:2504.19413}.

{[}3{]} Cormack, G. V., Clarke, C. L. A., \& Buttcher, S. (2009). Reciprocal rank fusion outperforms condorcet and individual rank learning methods. \emph{SIGIR}.

{[}4{]} Ebbinghaus, H. (1885). \emph{Uber das Gedachtnis}. Duncker \& Humblot.

{[}5{]} Ma, X., Wang, Y., \& Lin, J. (2024). Is reciprocal rank fusion all you need for hybrid retrieval? \emph{arXiv preprint}.

{[}6{]} Alqithami, S. (2025). Forgetful but Faithful: A Cognitive Memory Architecture and Benchmark for Privacy-Aware Generative Agents. \emph{arXiv:2512.12856}.

{[}7{]} Sarin, S., Singh, L., Sarmah, B., \& Mehta, D. (2025). Memoria: A scalable agentic memory framework for personalized conversational AI. \emph{arXiv:2512.12686}.

{[}8{]} Muennighoff, N., Tazi, N., Magne, L., \& Reimers, N. (2023). MTEB: Massive Text Embedding Benchmark. \emph{EACL}.

{[}9{]} Packer, C., et al.~(2023). MemGPT: Towards LLMs as operating systems. \emph{arXiv:2310.08560}.

{[}10{]} Dury, J. (2026). Predictive Associative Memory: Retrieval Beyond Similarity Through Temporal Co-occurrence. \emph{arXiv:2602.11322}.

{[}11{]} Santhanam, K., et al.~(2022). ColBERTv2: Effective and efficient retrieval via lightweight late interaction. \emph{NAACL}.

{[}12{]} Thakur, N., et al.~(2021). BEIR: A heterogeneous benchmark for zero-shot evaluation of information retrieval models. \emph{NeurIPS Datasets and Benchmarks}.

{[}13{]} Rasmussen, P., Paliychuk, P., Beauvais, T., \& Ryan, J. (2025). Zep: A temporal knowledge graph architecture for agent memory. \emph{arXiv:2501.13956}.

{[}14{]} Xu, W., Liang, Z., Mei, K., \& Gao, H. (2025). A-MEM: Agentic memory for LLM agents. \emph{arXiv:2502.12110}.

{[}15{]} Karpukhin, V., et al. (2020). Dense passage retrieval for open-domain question answering. \emph{EMNLP}.

{[}16{]} Xiong, L., et al. (2020). Approximate nearest neighbor negative contrastive learning for dense text retrieval. \emph{ICLR 2021}.

{[}17{]} Zhan, J., et al. (2021). Optimizing dense retrieval model training with hard negatives. \emph{SIGIR}.

{[}18{]} Lee, M., et al. (2024). NV-Retriever: Improving text embedding models with effective hard-negative mining. \emph{arXiv:2407.15831}.

{[}19{]} Chen, J., et al. (2024). BGE-M3: Multi-functionality, multi-linguality, and multi-granularity text embeddings through self-knowledge distillation. \emph{arXiv:2402.03216}.

\begin{center}\rule{0.5\linewidth}{0.5pt}\end{center}

\subsection{Acknowledgments}\label{acknowledgments}

vstash is built on sqlite-vec (Alex Garcia), FastEmbed (Qdrant), sentence-transformers and BAAI for embedding models, tree-sitter and parso for code-aware chunking, and SQLite/FTS5 for keyword retrieval. The BEIR evaluation suite (Thakur et al., 2021) provided the primary external benchmark. All design decisions, algorithm choices, experimental methodology, benchmark execution, interpretation of results, and the negative results documented in this paper are the author's own. The author reviewed and edited the final manuscript and takes full responsibility for its content. Development was assisted by Claude Opus 4 (Anthropic, 2025) for code generation, refactoring, and drafting of manuscript text. All experiments are reproducible from scripts in the project's \texttt{experiments/} directory.

\end{document}